\documentclass[%
 reprint,
superscriptaddress,
 amsmath,amssymb,
 aps,
]{revtex4-2}

\usepackage[utf8]{inputenc}
\usepackage[T1]{fontenc}
\usepackage{float}
\usepackage{siunitx}
\usepackage{relsize}
\usepackage{xspace}
\usepackage{algorithm}
\usepackage{algpseudocode}
\usepackage{graphicx}

\DeclareRobustCommand{\condor}{%
  Condor
}
\DeclareRobustCommand{\hawk}{%
  Hawk
}

\begin{document}

\title{Orientation Reconstruction of Proteins using Coulomb Explosions}

\author{Tomas Andr\'e}\thanks{These authors contributed equally to this work.}
\affiliation{%
 Department of Physics and Astronomy, Uppsala University, Box 516, SE-751 20 Uppsala, Sweden
}%

\author{Alfredo Bellisario}\thanks{These authors contributed equally to this work.}
\affiliation{%
 Department of Physics and Astronomy, Uppsala University, Box 516, SE-751 20 Uppsala, Sweden
}%

\author{Nicu\c{s}or T\^\i mneanu}\email{nicusor.timneanu@physics.uu.se}
\affiliation{%
 Department of Physics and Astronomy, Uppsala University, Box 516, SE-751 20 Uppsala, Sweden
}%

\author{Carl Caleman}\email{carl.caleman@physics.uu.se}
\affiliation{%
 Department of Physics and Astronomy, Uppsala University, Box 516, SE-751 20 Uppsala, Sweden
}%
 \affiliation{Center for Free-Electron Laser Science,
 Deutsches Elektronen-Synchrotron, \\ Notkestra\ss e 85, DE-22607 Hamburg, Germany}%

\begin{abstract}
We solve the orientation recovery of a tumbling protein in the gas phase from single-event measurements of the spatial positions of its ions after an X-ray laser–induced explosion. We simulate diffracted X-ray signal and ion dynamics under experimental conditions and compare our method to conventional orientation recovery in single-particle imaging with X-ray free-electron lasers using only diffraction data. We reconstruct 3D diffraction intensities using orientations recovered from the ion signatures and retrieve the electron density with established phase-retrieval algorithms. We test our orientation recovery procedure on 56 proteins ranging from 14 to 52 kDa (1800 to 6500 atoms), achieving roughly an angular error of around 5°. The resulting 3D electron-density reconstructions are compared to ground-truth volumes simulated at the same nominal resolution, and achieve the resolution at the edge of the detector in conditions similar to current single-particle imaging setups. We investigate the reconstruction quality and demonstrate that ion data can be used for reliable orientation recovery of particles in single-particle imaging, achieving orientation on par or better than currently used  recovery techniques. This work shows the potential of ion detection for retrieving additional information from the sample fragmentation, and boost single particle imaging with X-ray lasers in the cases where the diffraction signal is a limiting factor.

\end{abstract}
\raggedbottom

\maketitle

\thispagestyle{empty}
\section*{Introduction}

Proteins are central to life, carrying out a wide range of essential tasks such as enzymatic catalysis, signal transduction, and molecular transport. Because structure and function are tightly linked, resolving protein structures at high resolution is a key step toward understanding biological processes at the molecular scale. Crystallography~\cite{chua2022better} and cryo-electron microscopy~\cite{afonine2025macromolecular} are the workhorses of experimental structural biology~\cite{banari2025advancing}, allowing the study of hundreds of thousands of molecules with \AA ngstr\"{o}m resolution and, more recently, the development of AI-based computational tools such as AlphaFold \cite{jumper2021highly}, which has been proven to predict accurate protein structures directly from the amino-acid sequence. To image biologically relevant but rare conformations of single proteins, development of new methods is required.

\emph{Single Particle Imaging} (SPI)~\cite{neutze2000potential} with X-ray Free Electron Lasers (XFELs) allows one to study individual, non-crystalline samples in gas-phase, promising one day to achieve high resolution 3D reconstructions for weak scattering particles such as single proteins. In recent years, SPI has progressed considerably, with advances in sample delivery ~\cite{yenupuri2024helium} and analysis methods ~\cite{shen2021encryption}, opening a new way for the understanding of structural heterogeneity in nanoparticles ~\cite{shen2024resolving}. Though further development is required to fully realize its potential and to make it competitive with other imaging methods, SPI recently achieved a new milestone with the first observation of an X-ray diffraction pattern from a single protein structure ~\cite{ekeberg2024observation}.
\begin{figure}
    \centering
    \includegraphics[width=0.99\linewidth]{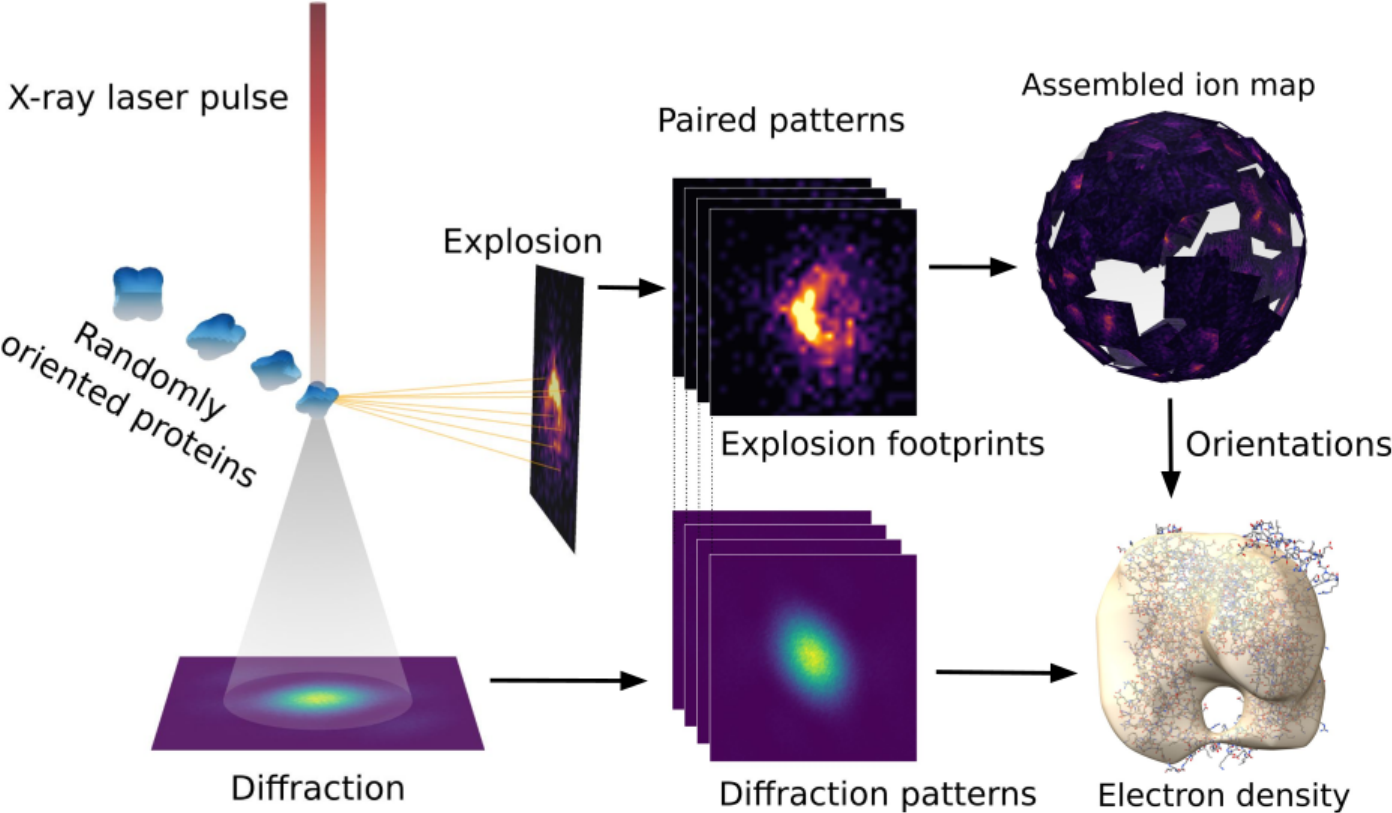}
    \caption{Randomly oriented proteins are hit by an X-ray laser to produce single-shot diffraction patterns and ion explosion footprints. Patterns and footprints corresponding to each concurrent event are paired; we find the  orientations of explosion footprints by mapping and assembling them on a spherical surface. These orientations are then applied on the diffraction patterns for reconstructing the 3D diffraction intensities space. Using phase retrieval algorithms we then reconstruct electron densities.}
    \label{fig:overview}
\end{figure}
As outlined early in an SPI road map~\cite{aquila2015roadmap}, one of the big challenges is the unknown orientation of the sample during delivery, as each sample can rotate freely while being delivered. So far, aligning molecules using laser-induced alignment~\cite{Alignment2014}, works only on small samples, outside the range of what can currently be imaged. Aligning large molecules such as proteins is still under active research, possibly using electric fields to align the dipole of the molecule during sample delivery~\cite{amin2025,Marklund2017orientation,SINELNIKOVA2021,agelii2025dipole}. Another approach is to use algorithms that can reconstruct 3D diffraction volumes from patterns with unknown orientation including Expand-Maximize-Compress (EMC)\cite{loh_reconstruction_2009} along with common arc~\cite{commonarc} and manifold embedding~\cite{moths2011bayesian}. In practice, orientation retrieval is performed only using EMC, which is the most successful algorithm developed so far. Despite its frequent use, EMC comes at the cost of no proven convergence, and while the effect of noise can be somewhat mitigated \cite{ayyer2019low} and symmetries of the sample can be taken into account to assist the convergence of reconstructions, high-resolution reconstructions necessitate ultimately hundreds of thousands of high-quality measurements \cite{you2025impact}, which is currently not feasible.

An outcome of single particle experiments which is often overlooked is ejected ions. In SPI, the X-ray pulse is intense enough to destroy the sample, but because the XFEL pulse is so short the interference can outrun the damage, and the pattern one measures is from the sample with the structure mostly intact at least theoretically. This is known as \textit{diffraction before destruction}~\cite{diffractB4destruct}. Collecting ions and photons simultaneously has already been performed experimentally, to assess the hit-rate of the beam on the sample~\cite{pietrini2018statistical} and to automate identification of hits~\cite{andreasson2014automated}. Although the explosion is governed by many-body Coulomb dynamics, where small changes in the initial configuration of the sample could produce disproportionately large changes in individual ion trajectories, the ensemble of ion positions should still preserve the underlying structural information. Recent studies have explored the structural information contained in the Coulomb explosion with different approaches, using COLTRIMS with momentum imaging~\cite{boll_x-ray_2022,ashrafi-belgabad_reconstructing_2024} as well as machine learning aided investigations~\cite{venkatachalam_exploiting_2025,li_generative_2026}. In previous works we showed that X-ray laser-induced explosions are reproducible~\cite{ostlin2018,EDS2024}; ion measurements can be used to classify protein conformations with differences as small as a twist in the $\beta$-hairpin loop for the protein of a viral capsid~\cite{andre2024}; we previously showed that the \textit{explosion footprints} generated can be classified using unsupervised machine-learning algorithms, including dimensionality reduction (e.g., PCA and t-SNE) and clustering, proving that in principle the ions maps contain information regarding the orientation of the sample~\cite{AndrePartial2025}. Drawing upon these earlier results, here we propose an algorithm that can find the relative orientation of the ion measurements of an exploding protein structure. 

The workflow, schematically illustrated in FIG.~\ref{fig:overview}, can be summarized as follows: (i) we simulate the Coulomb explosion and the diffraction signal from the same proteins in gas-phase at different orientations in concurrence. We then sample the simulations and pair the simulated explosion footprints with the corresponding diffraction patterns; (ii) we use the explosion footprints to determine the relative orientation of each measurement, then project and stitch the measurements onto a spherical surface to reconstruct a $4\pi$ sr  ion map; (iii) we use each diffraction pattern with their corresponding predicted orientations to assemble a 3D diffraction volume and retrieve the electron densities using phase retrieval. Ultimately, this paper addresses two scientific questions: \\(Q1) How well can we retrieve the orientation of a protein in gas-phase solely from the ejected ions of its Coulomb explosion induced by an XFEL beam? \\(Q2) How robust is the predicted orientation from ion maps when reconstructing the 3D structure from coherent diffraction?

\begin{figure}[h] 
    \centering
    \includegraphics[width=1\linewidth]{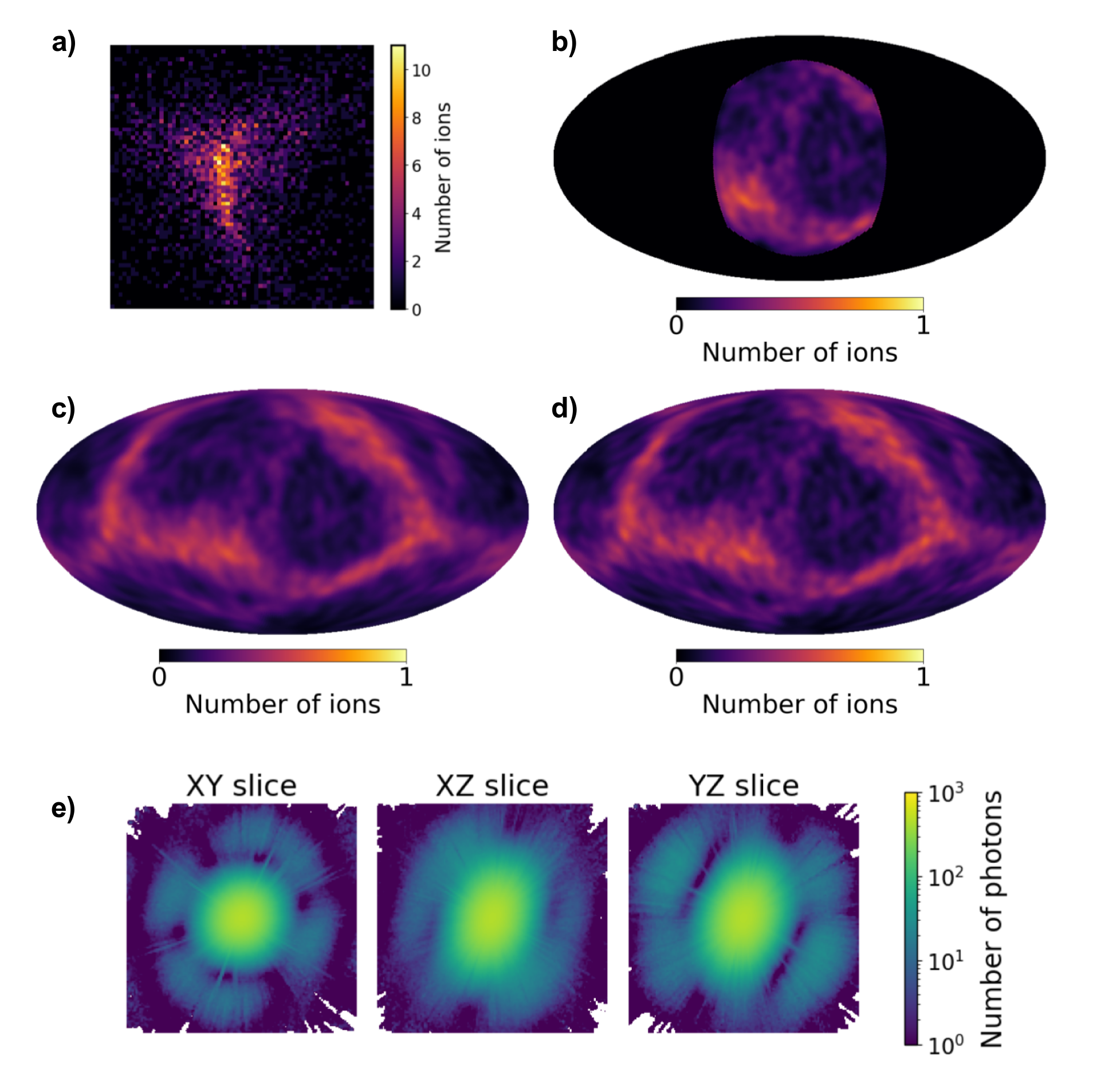} 
    \caption{Mapping and assembling the explosion and diffraction volume.
    We map the simulated Coulomb explosion footprints onto a sphere with radius equal to the distance from the interaction region. After finding the relative orientation of the sample from the ion maps we can retrieve the electron density in 3D using diffraction intensities. 
    a) Example of a binned simulated explosion footprint for an 80mm MCP located at 10~mm from the interaction region. b) Spherical projection of a single event, plotted using Mollweide view (non-integer intensities in the spherical projection arise from binning and Gaussian filtering). c) Spherical map of the full explosion estimated from the predicted orientations. d) Spherical map of the full explosion simulated from the sum of all the measurements when applying the exact orientations of the sample. e) Orthogonal slices (XY, XZ, YZ) through the assembled diffraction volume. Color bar indicates number of photons (log scale). 
    }
    \label{fig:data}
\end{figure}

\section*{Results}

\begin{figure}[h] 
    \centering
    \includegraphics[width=.95\linewidth]{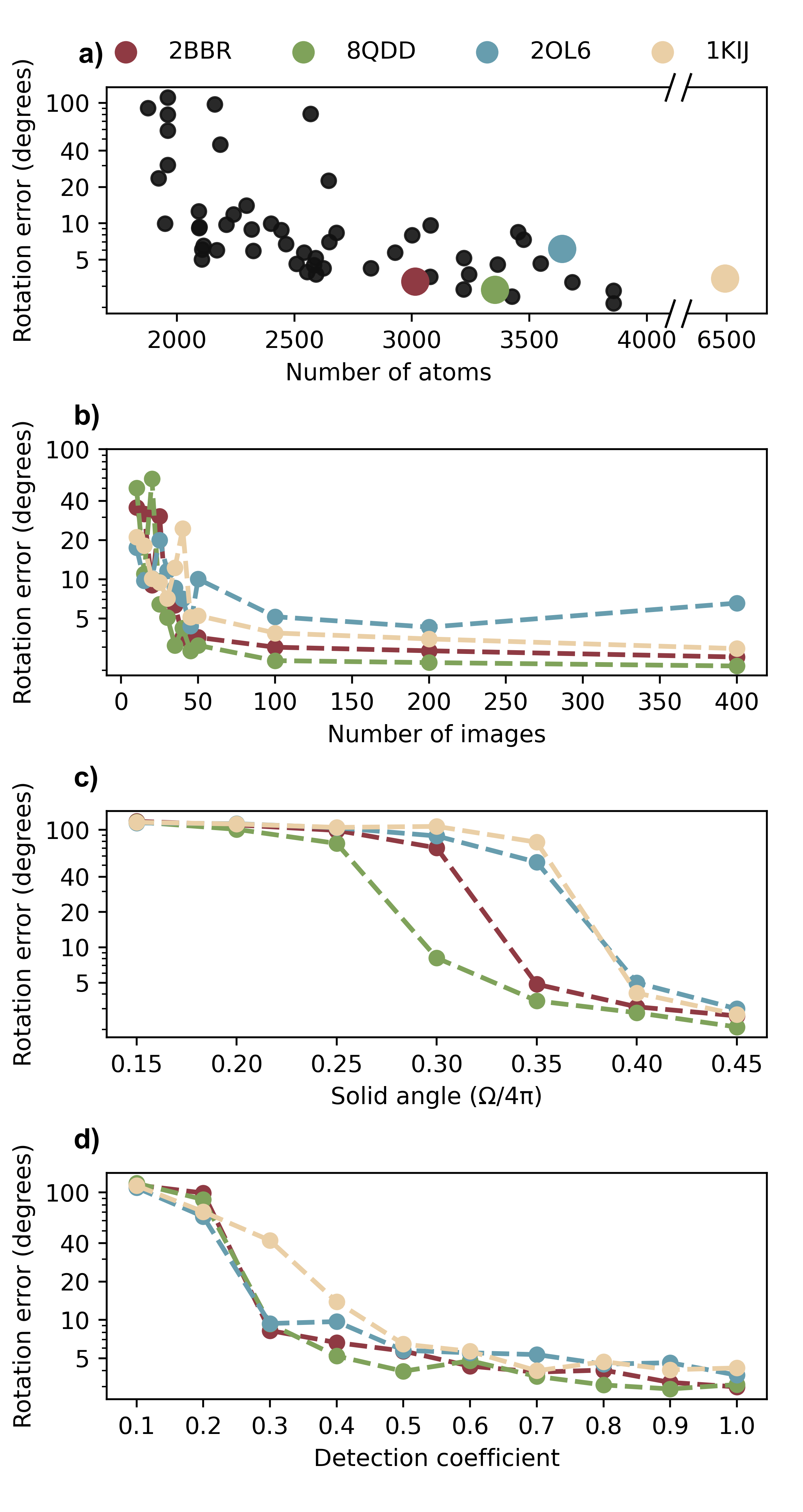}
    \caption{Benchmark of the orientation recovery algorithm against a) number of atoms, b) number of explosion footprints, c) solid angle covered by the detector, and d) detection coefficient (the chance that an ion that hit the detector is recorded). Rotation errors (degrees) are shown on the y-axis (log-scale), the benchmarks are carried out on the four representative proteins ordered by increasing molecular mass. In cases a),c),and d) the number of explosion footprints is 200.}
    \label{fig:benchmark}
\end{figure}

We simulated the X-ray laser-induced explosions (see Methods for details), on a set of 56 unique proteins 
and measure the ions on a virtual spatially resolved multi-channel plate detector with side-length 80~mm at 10~mm from the protein-laser interaction. This records a single-event time-integrated explosion footprint of the induced explosion. Each explosion footprint only contains a partial view of the ions, as not all ions are collected within the solid angle of the detector as shown in FIG.~\ref{fig:data}a. From the explosion footprint one can index the hit locations of the ions on the detector, and from these we construct spherical projection maps using the  Hierarchical Equal Area isoLatitude Pixelisation (HEALPix) algorithm for the pixelisation of a 2-sphere \cite{Gorski_2005,Zonca2019}, a method commonly used in cosmology and astronomy when working with spherical data.  
We display an example of an explosion footprint projected using HEALPix and displayed using a Mollweide projection in FIG.~\ref{fig:data}b. 

Using this procedure, we obtain the ion maps for random orientations in the laboratory frame. The next step is to find the relative orientation in the protein frame, thereby assembling a global \textit{ion map} which corresponds to the total \(4\pi\) sr solid angle of a complete sphere, covering the full angular distribution of the ions. To achieve this, we minimize the negative normalized cross-correlation between each pattern and a reference pattern selected from the set, and then iteratively refine the reference (see Methods for details). An example of a reconstructed ion map is shown in FIG.~\ref{fig:data}c and is compared to the ground truth assembled from known orientations as in FIG.~\ref{fig:data}d.

To benchmark our orientation retrieval procedure we explore four parameters: (a) sample size/mass; (b) the number of explosion footprints used; (c) solid angle covered by the detector, which can be controlled by changing the detector--sample distance; (d) the \textit{detection coefficient}, which simulates the efficiency of the detector in counting ions and corresponds to the ratio of detected ions and the total number of ions expected within the solid angle. The baseline parameters are 200 footprints, 1.0 detection coefficient, and a covered angle of \(0.42\)~str, as assumed in our simulations with a detector--sample distance of 10~mm for an 80~mm $\times$ 80~mm MCP. The dataset corresponds to simulations of 56 different protein structures. The results are all shown in FIG.~\ref{fig:benchmark} in the corresponding panels \ref{fig:benchmark}a-d, showing the best orientation achieved as a median error on the total number of iterations of our algorithm for the different benchmarks. Each plot varies only one parameter at a time, while the remaining baseline parameters are held fixed. There is a clear trend, see FIG.~\ref{fig:benchmark}a, connecting molecular mass and stable convergence of the algorithm. Running the algorithm for different proteins (FIG.~\ref{fig:benchmark}a) we see clearly that the bigger proteins are the easier to align; this could be attributed to ion trajectories being more well defined and reproducible as discussed in~\cite{EDS2024,AndrePartial2025}, and all proteins bigger than 3000 atoms successfully converge. 

We focus on four structures for tests (b), (c), and (d): viral CASP8 and FADD-like apoptosis regulator (2BBR~\cite{yang2005crystal}), outer surface protein A (2OL6~\cite{makabe2007beta}), mBaoJin green fluorescent protein (8QDD~\cite{zhang2024bright}), and DNA gyrase subunit B (1KIJ~\cite{LAMOUR200218947}).  The choice of these structures falls on four proteins that allow us to probe the performance of the algorithm across a wide molecular mass range to be sufficient for a stable convergence while also identifying molecules with different structural features - small and compact, elongated, barrel shaped, and dimer - hence with different signature in the global ion map.  When varying the number of images, see FIG.~\ref{fig:benchmark}b, we quickly reach successful reconstructions, although quite unstable until 100 images, for small amount of images we might not completely sample the sphere and have few overlapping images, making reconstruction difficult. For 100 and more images the error remains low and stable. In the plot we show as a baseline the lowest error achieved with EMC on the same four proteins with 400 patterns; our procedure outperforms orientation retrieval on diffraction patterns, reaching a lower error and requiring less data. 

FIG.~\ref{fig:benchmark}c shows that the most compact molecules, 2BBR and 8QDD, are also the ones that require a small coverage of solid angle for reliable convergence; on the axis we scale the solid angle by a factor of \(\frac{1}{4\pi}\) to clarify what ratio of the sphere needs to be covered. From the solid angle, we see that we need a setup such that around 35\% to 40\% of the solid angle is covered, which could be achieved by combinations of detector--sample distance and detector sizes of currently used MCP detectors. For the detection coefficient, see FIG.~\ref{fig:benchmark}d,  the error remains low up until we only detect around 30\% of the ions. The robustness down to \(\sim30\)\% detection suggests that orientation information is encoded in global anisotropic features of the explosion footprint rather than in individual ion trajectories. This is further confirmation that the orientation information is kept in global features, if each sample does not cover a large enough patch of the sphere, these global distribution features can no longer be correlated between patches, the explosion footprints, of the total ion map. 

\begin{figure}[h] 
    \centering
\includegraphics[width=1\linewidth]{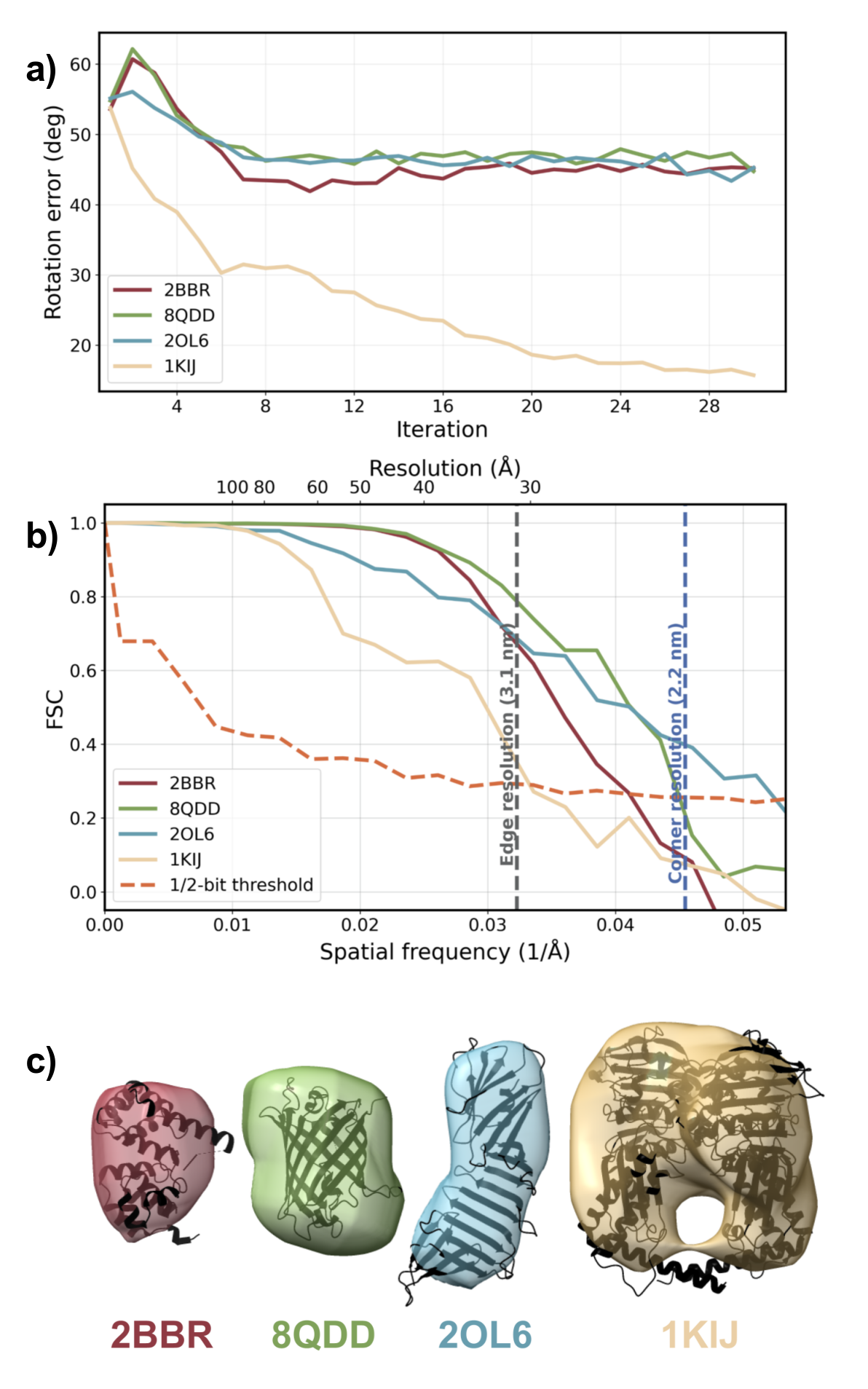}
    \caption{SPI data analysis using diffraction patterns. a) Orientation error as a function of EMC iteration using 400 diffraction patterns for four representative proteins. Orientation errors are reported modulo $D_2$ (crystallographic $222$). b) Fourier shell correlation (FSC) plot. The dashed orange line indicates the 1/2-bit FSC threshold, and the vertical dashed lines mark respectively the edge and the corner resolutions (3.1 nm and 2.2 nm). The achieved reconstruction resolutions, determined from the 1/2-bit FSC criterion, are $24.3\,\text{\AA}$ for 2BBR, $19.0\,\text{\AA}$ for 2OL6, $22.2\,\text{\AA}$ for 8QDD, and $30.0\,\text{\AA}$ for 1KIJ. All reconstructions achieve a resolution better than the edge resolution limit of $30.1\,\text{\AA}$.
    c) Representative reconstructed 3D volumes with embedded protein models. The density map was visualized in ChimeraX using an isosurface threshold of 15\% of the map intensity range~\cite{meng2023ucsf}.}
    \label{fig:diffraction}
\end{figure}

Using the same X-ray exposure conditions 
for the ion explosion simulations, we simulate the corresponding diffraction patterns for our benchmark protein structures. We simulate a detector of 1024 by 1024 pixels, 200~$\mu$m pixels size, downscaled by a factor 8, and positioned at 50 cm from the interaction region, which limits the best achievable resolution to 2.2 nm calculated at the corner of the detector, comparable to experimental setups currently achievable for SPI at XFEL facilities~\cite{allahgholi2019megapixels,donato2017first,meyer2022sqs,mancuso2013technical}. In FIG.~\ref{fig:diffraction}a, we show the best performance for each benchmark PDB structure of 10 EMC instances. Given the small size of the proteins and the low resolution, we can assume $D_2$ symmetry of the sample, which further lowers the final rotational error of EMC. Under the given experimental setup and parameters, we note that even for 400 patterns, EMC does not reach below 15\(^\circ\) error. As a rule of thumb, orientation recovery is successful if the median angular error is less than 10\(^\circ\), consistent with thresholds adopted in other EMC-based reconstruction studies~\cite{august2024,wollter2024coherent,wollter2024phase}. It is important to note that the smallest theoretical angular error of EMC depends on the sampling resolution parameter $n$, which sets how finely the space of quaternions is discretized~\cite{loh_reconstruction_2009}: increasing $n$ reduces the angular spacing, but also increases the number of sampled orientations and thus the computational cost which scales approximately as $n^3$. For this study, we set $n = 20$, which corresponds to a theoretical smallest error of $2.7^\circ$. In contrast, the ion detector has an angular resolution of \(3.8^\circ\) for the central pixels and \(1.8^\circ\) on the outer pixels. This results in the ion measurements being sampled more finely than EMC, and hence the angular resolution in this comparison is also limited by the EMC discretization. The angular spacing affects the fluctuations in the convergence of EMC~\cite{loh_reconstruction_2009,wollter2024phase} and limits the best achievable resolution when using phase retrieval. With our results, in FIG.~\ref{fig:benchmark}a and~\ref{fig:diffraction}a, we show that it is possible to achieve low errors, comparable to the theoretical optimum with EMC, even under conditions where EMC would otherwise fail to converge; without a converged EMC, the diffraction intensities volume cannot be reliably phased, and accurate real-space electron-density reconstruction is therefore not possible. 

FIG.~\ref{fig:diffraction}b shows the four benchmark proteins and their electron densities obtained after phase retrieval on the assembled 3D diffraction intensities using orientations as predicted in FIG.~\ref{fig:benchmark}a. We then use the Fourier Shell Correlation (FSC)~\cite{liao2010definition} criterion to estimate the resolution of our reconstructions, as in FIG.~\ref{fig:diffraction}c. With ground truth available from the PDB files, we compare directly to the electron densities estimated at the corner resolution with ChimeraX~\cite{pettersen2021ucsf} rather than using half-sets. The resolution can be estimated with the half-bit criterion ~\cite{van2005fourier}, but, in this setting, our reconstructions are very small volumes and the FSC remains mostly flat and close to 1 even when close to the edge of the detector. It might be that there is little independent high-frequency noise to decorrelate the maps and this indicates that we are achieving the best reconstructions we can with the current detector setup.

\section*{Discussion and Conclusion}

With regard to the questions posed in the introduction:

\begin{enumerate}
  \renewcommand{\labelenumi}{(Q\arabic{enumi})}
  \item we can find the orientation of the sample from ion maps down to \(\sim5^\circ\), and we achieve the best possible resolution in electron density reconstruction with the given diffraction parameters. 
  \item we require fewer ion measurements compared to diffraction measurements, down to 50--100 ion maps for reliable convergence, at the expense of collecting at least one third of the entire explosion; moreover, we notice a soft lower limit of \(\sim\)2500 atoms on the sample size required for convergence;
\end{enumerate}
This study has some potential limitations, for example in the generation of explosion footprints, the effect of a realistic ion background or the shot-to-shot intensity variations of an exposure, are not included. On the explosion model side, we use one software for simulating dynamics and rely on the explosion reproducibility, however other software could also be tested. Within the model we are limited to small samples under the assumption that free electrons leave the sample, while larger samples would capture the electrons and have a more isotropic explosion. On the imaging side, we have assumed a fixed detector geometry to match the conditions at the SQS instrument at the European XFEL. 
We have considered only one conformation of proteins, under the assumption that conformations can be prepared or separated experimentally using mass spectroscopy.
A further limitation to note is the many parameters in the orientation retrieval method that are fixed for the whole data set, some of them include level of explosion pattern smoothing, number of bins in the HEALPix projection, and pairwise overlap/similarity score threshold to accept and reject an orientation. These parameters can be tweaked for better results for smaller mass ranges. For example, we successfully retrieved the orientation of the small protein ubiquitin, PDB-ID 1UBI~\cite{ramage1994synthetic} ($\sim1200$ atoms, counting hydrogens).

Heterogeneity classification and conformational studies are still open problems and active research areas in SPI. While the latest developments push towards smaller samples and higher resolution, it will be become increasingly important to sort and classify heterogeneous datasets. As shown in publications related to this work~\cite{ostlin2018,andre2024,AndrePartial2025}, the spatial position of ions \emph{post-destruction} in SPI contains sufficient information to disentangle small conformational changes. Rare conformational states would also less frequently be probed by diffraction, leading to less data, which can be a bottleneck for current SPI data analysis. In this work, assuming static and non-heterogeneous structures, we provide a pathway to orientation retrieval using explosion fingerprints and very scarce measurements. This work can help accessing structures that wouldn't otherwise be possible to analyse. The fact that explosion footprints have been shown to be unique for each protein ~\cite{andre2024} hints at the possibility to map explosion footprints to structure. This combined with the results described here, provides a strong scientific case for ion collection during SPI. A link between structures and explosion maps would allow structural features to be determined with  tabletop lasers, completely omitting the diffraction.



\section*{Methods}
\subsection*{A. Ion trajectories}
\subsubsection*{Simulation setup}

All ionization and Coulomb explosion simulations are performed with \textsc{MolDStruct}~\cite{dawod_moldstruct_2024}, a Monte Carlo/Molecular Dynamics framework implemented in \textsc{Gromacs}~4.5.6, using the CHARMM36 force field. Each protein structure is first energy-minimized (steepest descent) and equilibrated in vacuum at 300~K with a 1~fs timestep while keeping its orientation fixed. From the equilibrated trajectory we extract initial configurations and run 100 ionization simulations per protein. For the cases where we use more than 100 explosion footprints for the analysis, we sample the initial 100 from different orientations. The ionization model includes photo-ionization, fluorescence, and Auger--Meitner decay, and assumes that all electrons escape, consistent with system sizes considered here. Particle injection speeds (10$^{1}$--10$^{2}$~m/s) are negligible compared to the ion velocities during Coulomb explosion (10$^{4}$--10$^{5}$~m/s). The Coulomb explosion dynamics are propagated with a timestep of 1~as. The XFEL pulse is modeled as a Gaussian with peak at $t=20$~fs, a FWHM of 10~fs, photon energy of 2~keV, and fluence of $5\times10^{6}$~photons/nm$^{2}$, consistent with conditions at the SQS instrument of the European XFEL \cite{meyer2022sqs}. Each simulation runs for 250~fs, sufficient for the system to reach the kinetic-energy–dominated regime where ion--ion repulsion becomes negligible.

\subsubsection*{Orientation retrieval algorithm}

An explosion footprint \(I\) represents the spatial distribution of ion intensities recorded for a single Coulomb explosion event. \(I\) is represented by ion directions projected onto a partial sphere using the HEALPix projection CCCite.  Given \(N\) such footprints \(I_i\), each captured under an unknown orientation \(R_i^{\mathrm{true}}\), our goal is to determine rotations \(\hat{R}_i\) that align all footprints into a common frame. One footprint is selected as an initial model \(M_0\), used as the reference during the first iteration. Each footprint \(I_i\) is expanded into \(n\) trial orientations
\[
I_i^{\,j} = G_j I_i,
\]
where the \(G_j\) are rotations generated via Sobol-based quaternion sampling~\cite{SOBOL1967,shoemake1992uniform}, yielding a quasi-uniform coverage of \(SO(3)\). We define applying an orientation $R$ to a HEALPix map $I_i$ as
\[
(RI)(\mathbf{n}) = I(R^{-1}\mathbf{n}),
\]
meaning that the rotation acts on the coordinate frame of the sphere rather
than on the map values themselves.

To evaluate the match between a rotated footprint and the model, we use the zero-mean normalized cross-correlation (ZNCC):
\begin{equation}
\mathrm{ZNCC}(a,b)
=
\frac{\displaystyle\sum_{k=1}^{N}
(a_k - \bar{a})(b_k - \bar{b})}
{\sqrt{\displaystyle\sum_{k=1}^{N}(a_k - \bar{a})^2}\,
 \sqrt{\displaystyle\sum_{k=1}^{N}(b_k - \bar{b})^2}},
\end{equation}
with the means
\begin{equation}
\bar{a}
= \frac{1}{N}\sum_{k=1}^{N} a_k,
\qquad
\bar{b}
= \frac{1}{N}\sum_{k=1}^{N} b_k.
\end{equation}

For each trial rotation we compute
\begin{equation}
\mathrm{ZNCC}\!\left(M_0,\; I_i^{\,j}\right),
\end{equation}
using only the pixels where both maps are defined.
This allows us to rank the trial orientations of footprint \(I_i\) by their agreement with the reference.

The top \(k\) candidates undergo a local refinement step: starting from each coarse rotation \(G_j\), we perform a Powell optimization in Euler-angle space to minimize
\begin{equation}
\Delta G^\star = \arg\min_{\Delta G}\;
    \big[-\mathrm{ZNCC}(M_0,\, \Delta G\, G_j I_i)\big],
\end{equation}
yielding an optimized alignment rotation \(\hat{R}_i = \Delta G^\star\, G_j\).

After processing all footprints, the aligned maps \(\hat{R}_i I_i\) with ZNCC scores above a threshold are averaged to form an updated model \(M_1\). This procedure is repeated iteratively, replacing \(M\) with the most recent model until convergence.

\begin{algorithm}[H]
\caption{Orientation Retrieval}
\label{alg:refinement}
\begin{algorithmic}[1]

\Require Patterns $\{I_i\}_{i=1}^N$
\State Select one pattern as initial model $M_0$
\State Generate Sobol-based quaternions $\{G_j\}_{j=1}^n$ for quasi-uniform coverage of $SO(3)$

\For{$i = 1$ to $N$}
    \State Construct trial orientations $I_i^{\,j} = G_j I_i$ for all $j$
    \State Compute ZNCC scores
    \[
        s_{i,j} = \mathrm{ZNCC}(M_0, I_i^{\,j})
    \]
    \State Select top-$k$ orientations according to $s_{i,j}$
    \For{each selected $G_j$}
        \State Refine rotation via Powell optimization:
        \[
           \Delta G^\star = \arg\min_{\Delta G}
                \left[-\,\mathrm{ZNCC}(M_0, \Delta G\, G_j\, I_i)\right]
        \]
    \EndFor
    \State Set recovered orientation $\hat{R}_i = \Delta G^\star\, G_j$ with the best refined score
\EndFor


\State Form updated model by averaging aligned footprints:
\[
    M_1 = \frac{1}{|N'|}\sum_{i \in N'} \hat{R}_i I_i,
\]
where $N' = \{ i \mid \mathrm{ZNCC}(M_0, \hat{R}_i I_i) > \tau \}$

\State Replace $M_0 \leftarrow M_1$ and repeat until convergence

\end{algorithmic}
\end{algorithm}

\subsection*{B. X-ray diffraction}

\subsubsection*{Diffraction data simulations}

We simulate diffraction patterns using \condor \cite{hantke2016condor} to compare our results with orientation recovery in SPI experiments. 
Simulations have been performed in accordance to beam parameters used for the explosion simulations with Molecular Dynamics; specifically, using a photon energy of 2 keV, a pulse energy of 3\(\times10^{-1}\)J, and a focus diameter of \(200\,\mathrm{nm}\). We place a 1024 pixels detector at \(0.5\,\mathrm{m}\), with each pixel  \(200\,\mu\mathrm{m}\) long, and we downsample it by a factor 8, assuming a flat Ewald sphere. Let $F(\mathbf{q})$ denote the complex Fourier amplitudes of the object, and let 
\[
\mathcal{I}_i(\mathbf{q}) = |F(R_i^{\mathrm{true}}\mathbf{q})|^2
\]
be the $i$-th simulated diffraction pattern generated under a known orientation $R_i^{\mathrm{true}}$.

The orientations recovered from the ion maps, denoted $\hat{R}_i$, are used to assemble the 3D Fourier intensity model
\[
W_0(\mathbf{q}) = \frac{1}{N} \sum_{i=1}^N \mathcal{I}_i(\hat{R}_i \mathbf{q}) .
\]
\newline

\subsubsection*{Orientation retrieval of the diffraction patterns} 

We compare the volume $W_0$ with standard diffraction-based recovery, by applying the EMC algorithm~\cite{loh_reconstruction_2009} directly to the diffraction patterns $\mathcal{I}_i$. Let 
\[
P_{i}(R) = \Pr(R \mid \mathcal{I}_i, W)
\]
be the posterior orientation probability assigned by EMC for pattern $i$ given a 3D model $W$. Each EMC reconstruction is initialized with a random probability distribution over $SO(3)$, and iterated until convergence.  Since EMC performance varies with initialization, each parameter set is run multiple times with independent random seeds, and the best-scoring reconstruction (e.g., by relative rotational error over known orientations) over 10 instances is retained.

\subsubsection*{Electron density reconstructions} 
The final electron density \(\rho(\mathbf{r})\) is recovered by iterative phase retrieval, alternating between Relaxed Averaged Alternating Reflections (RAAR) ~\cite{luke2004relaxed} and Error Reduction (ER)~\cite{fienup1978reconstruction}. Phase retrieval is performed on \(W_0\) using \hawk~\cite{maia2010hawk} with GPU acceleration. Let \(\mathcal{P}_m\) denote the Fourier-modulus (measured amplitude) projector, \(\mathcal{P}_S\) the real-space support projector, and \(I\) the identity operator, and define the corresponding reflectors
\[
R_m = 2\mathcal{P}_m - I, \qquad R_S = 2\mathcal{P}_S - I .
\]
The RAAR update is
\[
\rho_{k+1}=
\left[
\frac{\beta}{2}\left(R_S R_m + I\right) + (1-\beta)\mathcal{P}_m
\right]\rho_k ,
\qquad \beta=0.9, 
\]
and the ER update is
\[
\rho_{k+1}=\mathcal{P}_S\,\mathcal{P}_m\,\rho_k .
\]

Each reconstruction uses the following iteration schedule:
\[
10{,}000~\text{RAAR iterations}
\;\rightarrow\;
2{,}500~\text{ER iterations},
\]
for a total of \(12{,}500\) iterations. We initialize the reconstruction with a spherical support \(S_0\) of radius \(20\) voxels and update it using the Shrinkwrap algorithm~\cite{marchesini2003x} every 100 iterations. During each Shrinkwrap update, the current density is convolved with a Gaussian kernel whose standard deviation decreases from $\sigma = 2$ voxels at the beginning of the reconstruction to $\sigma = 1$ voxels at the final iteration.

\bibliography{references}

\section*{Acknowledgements}
Project grants from the Swedish Research Council (2018-00740, 2019-03935, 2023-03900) are acknowledged, and the Helmholtz Association through the Center for Free-Electron Laser Science at DESY. 
CC acknowledges support from a Röntgen Ångström Cluster grant provided by the Swedish Research Council and the Bundesministerium für Bildung und Forschung (2021-05988). The computations were enabled by resources in projects NAISS 2025/5-375, 2025/22-229, 2026/4-258 provided by the National Academic Infrastructure for Supercomputing in Sweden (NAISS), funded by the Swedish Research Council through grant agreement no. 2022-06725. Computations related to the electron density analysis were performed on the Davinci computer cluster at the Laboratory of Molecular Biophysics, Uppsala University. Molecular graphics performed with UCSF ChimeraX, developed by the Resource for Biocomputing, Visualization, and Informatics at the University of California, San Francisco, with support from National Institutes of Health R01-GM129325 and the Office of Cyber Infrastructure and Computational Biology, National Institute of Allergy and Infectious Diseases.
\\

\section*{Author contributions statement}
T.A.: Conceptualization, Design, Simulations, Analysis, Interpretation, Writing - first draft \& editing, A.B.: Conceptualization, Design, Simulations, Analysis, Interpretation, Writing - first draft \& editing, N.T.: Conceptualization, Design,  Interpretation, Writing, C.C.:  Conceptualization, Design,  Interpretation, Writing. All authors reviewed the manuscript.

\end{document}